\documentclass[10pt,letterpaper]{article}
\renewcommand{\thesubsection}{\Alph{subsection}}
\usepackage{enumitem}
\usepackage{amsmath}
\usepackage{amsfonts}
\usepackage[T1]{fontenc}
\usepackage{amssymb}
\usepackage{color}
\usepackage{graphicx}
\usepackage{epstopdf}
\usepackage{bbold}
\usepackage{comment}
\usepackage{color}
\usepackage{soul}
\usepackage[abs]{overpic}
\usepackage{amsmath,amsfonts,amssymb}
\usepackage{lipsum}
\usepackage{hyperref}
\usepackage{setspace}
\usepackage{siunitx}
\usepackage{afterpage}
\usepackage{braket}
\usepackage{bm}
\usepackage[bottom]{footmisc}
\DeclareSIUnit\torr{Torr}
\DeclareSIUnit\gauss{G}

\newcommand{\beq}{\begin{equation}}
\newcommand{\eeq}{\end{equation}}

\newcommand{\cm}{\mbox{ cm}}

\newcommand{\Rb}{\ensuremath{^{87}}Rb }

\begin{document}

% title goes here:
\begin{flushleft}
{\Large
\textbf\newline{Structured beams invariant to coherent diffusion}
}
\newline
% authors go here:
\\
Slava Smartsev\textsuperscript{*},
Ronen Chriki,
David Eger,
Ofer Firstenberg,
and Nir Davidson
\\
\bigskip
Department of Physics of Complex Systems, Weizmann Institute of Science, Rehovot 7610001, Israel
\\
\bigskip
* slava.smartsev@weizmann.ac.il

\end{flushleft}

%\title{Structured beams invariant to coherent diffusion}
%\author{Slava Smartsev\authormark{*}, Ronen Chriki, David Eger, Ofer Firstenberg, and Nir Davidson}

%\address{Department of Physics of Complex Systems, Weizmann Institute of Science, Rehovot 7610001, Israel}

%\email{\authormark{*}slava.smartsev@weizmann.ac.il} %% email address is required

%%%%%%%%%%%%%%%%%%% abstract %%%%%%%%%%%%%%%%

\begin{abstract}
Bessel beams are renowned members of a wide family of non-diffracting (propagation-invariant) fields. We report on experiments showing that non-diffracting fields are also immune to diffusion. We map the phase and magnitude of structured laser fields onto the spatial coherence between two internal states of warm atoms undergoing diffusion. We measure the field after a controllable, effective, diffusion time by continuously generating light from the spatial coherence. The coherent diffusion of Bessel-Gaussian fields and more intricate, non-diffracting fields is quantitatively analyzed and directly compared to that of diffracting fields. To elucidate the origin of diffusion invariance, we show results for non-diffracting fields whose phase pattern we flatten.
\end{abstract}

%%%%%%%%%%%%%%%%%%%%%%%%%%body%%%%%%%%%%%%%%%%%%%%%%%%%%
\section{Introduction}
Diffusion is a fundamental, well-studied physical phenomenon. For real-valued scalar fields, such as pressure, heat, and concentration, spatial diffusion acts to suppress the field gradients and thus operates as a low-pass filter in the spatial domain. As a result, diffusion smooths or broadens local features of real-values fields. 

Complex-valued scalar fields, which have both magnitude and phase, may also be subjected to diffusion. Diffusion of complex fields is often referred to as \emph{coherent} diffusion and encountered in polarized spin ensembles, including nuclear magnetization in NMR \cite{TorreyPR1956,NMRBrownianRMP2007}, spin-polarized atoms in vapor \cite{Skalla_1997,ShukerImaging2007,HeinzePRL2013, arxiv_shaham2020}, and electronic or exciton spins in spintronics systems, such as metals, ferromagnets, and semiconductors \cite{KaplanPR1959,SpintronicsRMP2004,HauNature1999,JansonNature2009}. When such complex fields diffuse, their phase plays an important role and can lead to strikingly different evolution than simple spatial spreading and smoothing. For example, diffusion of stored and slow light in a warm atomic vapor \cite{FirstenbergRMP2013} have been employed to demonstrate spatial contraction of complex fields \cite{FirstenbergPRL2010}, topological protection of optical vortices \cite{FirstenbergPRL2007}, self-similar expansion of so-called elegant Gaussian modes \cite{FirstenbergPRL2010,YankelevOL2013}, and diffusion-induced diffraction that can modify, eliminate, and even reverse the optical paraxial diffraction \cite{FirstenbergNP2009}.

The above examples illustrate similarities between coherent diffusion and optical paraxial diffraction, which can be traced back to the equations describing these distinct phenomena. Let $\psi(\mathbf{r},z)$ and $\psi(\mathbf{r},t)$ be complex-valued fields, where $\mathbf{r}=(x,y)$ are the transverse spatial coordinates, and consider the paraxial monochromatic Helmholtz equation and Fick's second law of diffusion in real space, 
\beq
\Big(\frac{\partial}{\partial z}-i\frac{\lambda}{4\pi}\nabla^2_\bot\Big)\psi(\mathbf{r},z)=0,
~~~~\Big(\frac{\partial}{\partial t}-D\nabla^2_\bot\Big)\psi(\mathbf{r},t)=0. 
\eeq
The diffusion constant $D$ is real, whereas the diffraction constant $i\lambda/4\pi$ ($\lambda$ is the optical wavelength) is imaginary. Both propagators are more naturally described in the spatial frequency space, acting as products on the Fourier transform of the fields $\tilde{\psi}(\mathbf{q})$, where $\mathbf{q}$ are the transverse spatial frequencies. Paraxial diffraction is caused by spatial dispersion, \emph{i.e.}, from dephasing between the spatial frequencies \cite{Goodman2005}; the diffraction propagator multiples $\tilde{\psi}(\mathbf{q})$ by the phase factor $\exp(-i\frac{\lambda}{4\pi}q^2z)$ while maintaining the magnitude. As no dissipation occurs, diffraction can in principle be perfectly reversed. Conversely, the diffusion propagator reduces the amplitude of $\tilde{\psi}(\mathbf{q})$ by $\exp(-Dq^2 t)$ causing irreversible decay of the amplitude and the correlations of the initial field, which is faster for higher spatial frequencies \cite{Chriki_2019_Optica}.

Optical beams whose spatial frequencies lay in Fourier space on a ring of zero width, namely Bessel beams \cite{Durnin_diff_free} and all of their super-positions \cite{uno1995speckle, turunen1991propagation}, conserve their transverse profile when diffracting. This follows from the diffraction propagator depending only on $q=|\mathbf{q}|$. Such beams are non-physical, for example because they have diverging moments. However, their physically relevant approximations, such as Gauss-Bessel beams obtained by multiplying a Bessel beam by a large Gaussian envelope, have been shown to be invariant to diffraction over finite but arbitrarily-long propagation distances \cite{Bessel_Gauss}. We review the proof of diffraction invariance of Bessel beams in Appendix \ref{appendix:bessel_invariance}.
 
In this work, we show by similar arguments that Bessel beams and their superpositions are spatially invariant also to coherent diffusion and that their Gauss-Bessel approximations are invariant to coherent diffusion over long times. We experimentally demonstrate this striking invariance to diffusion using structured light imprinted on and read out of diffusing atomic vapor. For comparison, we repeat the experiment with fields which are not invariant to diffraction and observe significant changes in their structure as they diffuse. Finally, we show that real-valued fields that have the intensity profile of a Bessel beam are not invariant to diffusion, thereby stressing the importance of the complex phase in diffusion invariance.

\section{Setup}
For the experimental study of diffusion of structured complex fields, we exploit electromagnetically induced transparency (EIT) within a unique four-wave mixing scheme studied in Ref.~\cite{smartsev2017continuous}. The spatially-shaped `probe' beam is imprinted onto warm atoms, and a `signal' beam is continuously retrieved from them and measured. The atoms diffuse during this process, and the retrieved signal conveys their diffusion evolution. The effective diffusion time is equal to the group delay $\tau$ due to EIT in the medium \cite{smartsev2017continuous,Chriki_2019_Optica}.

\begin{figure}[t]
		\centering
		\includegraphics[width=10cm]{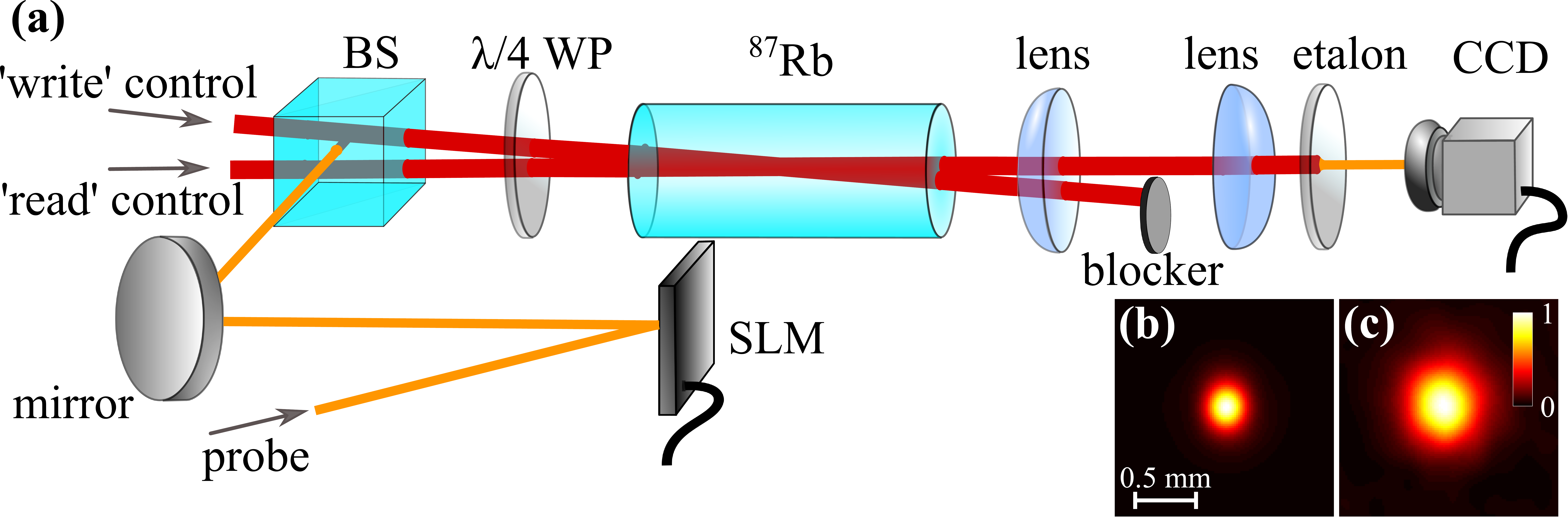}
		\caption{\label{fig:setup} \small {Simplified setup and representative results. \textbf{(a)} Experimental arrangement for realizing the diffusion of structured fields (BS - beam splitter; SLM - spatial light modulator; $\lambda/4$ WP - quarter-wave retarder; CCD - camera). \textbf{(b)} Measured image of a probe field (along the direction of the `write' control) that is imprinted onto the atomic coherence. \textbf{(c)} Measured normalized intensity profile of the signal field (along the direction of the `read' control) retrieved from the atomic coherence after an effective diffusion time of $\tau=37~\mu\mathrm{s}$. The colorbar shown in \textbf{c} is the same throughout the paper.}}
\end{figure}
 
Our experimental arrangement is illustrated in Fig.~\ref{fig:setup}(a). We use a vapor cell with \Rb atoms at $65^\circ\mathrm{C}$, which diffuse in \SI{10}{\torr} of $\text{N}_2$ buffer gas, rendering a diffusion coefficient of $D=\SI[separate-uncertainty = true, multi-part-units=single]{9.7(5)}{\cm\squared\per\second}$. The cell is continuously illuminated by spatially overlapping `write' and `read' control beams, which are separated by a small angle, and by a weak `probe' beam propagating along the `write' control. Consequently, a fourth beam, denoted as `signal', is generated in a four-wave mixing process along the direction of the `read' control. We set the optical frequencies of the probe and control beams to match the \Rb D1 transition from, respectively, the lower and upper hyperfine states in the ground level. The incoming probe field $E_\text{in}(\mathbf{r})$ is shaped using a spatial light modulator (SLM). The outgoing signal $E_\text{s}(\mathbf{r})$ is separated from the `read' control by a Fabry-P\'{e}rot etalon and imaged onto a camera, and we use digital Fourier filtering to improve the signal-to-noise ratio. Further details on the experimental arrangement are given in Appendix~\ref{appendix:exp_setup}.

Representative recorded images of an incoming probe $E_\text{in}(\mathbf{r})$ shaped as a Gaussian and the corresponding retrieved signal $E_\text{s}(\mathbf{r})$ are shown in Figs.~\ref{fig:setup}(b) and \ref{fig:setup}(c). As shown in details in Appendix~\ref{appendix:EIT_delay}, $E_\text{s}(\mathbf{r})$ is effectively delayed with respect to $E_\text{in}(\mathbf{r})$ by a duration $\tau$, thus undergoing and conveying the diffusion evolution. The diffusion time $\tau$ depends on the intensity and frequency of the control fields [see Eq.~(\ref{eq:diffusion_time2})]; In the experiment, we vary $\tau$ by tuning the control frequency.

\section{Results}

\begin{figure}[t]
		\centering
		\includegraphics[width=10cm]{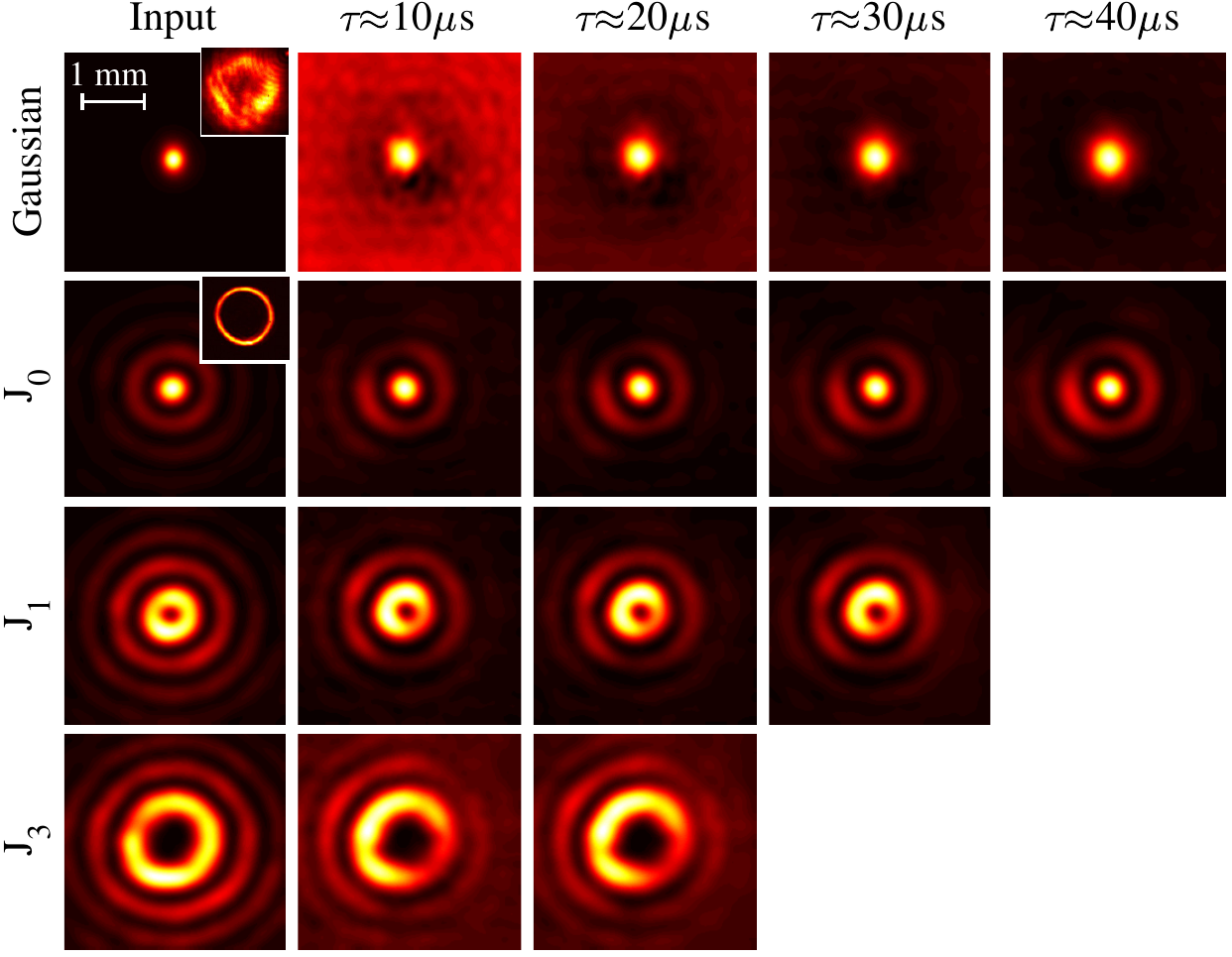}
		\caption{\label{fig:Beams_table} \small {Invariance of Bessel-Gauss beams to coherent diffusion. Shown are normalized measured intensity profiles. The left column presents the input probe, which sets the initial profile of the atomic coherence field. Insets present the same probe measured in the far-field, which is approximately a Gaussian for the Gaussian beam and a narrow ring for the Bessel-Gauss beam. The columns labeled by $\tau\approx10$, $20$ ,$30$, $40~\mu\mathrm{s}$ show the generated signal after the effective diffusion time $\tau$. While the standard Gaussian beam expands (top row) the Bessel-Gauss beams $J_0$, $J_1$, $J_3$ are nearly invariant to diffusion (second, third, and fourth rows).}}
	\end{figure}
	
\begin{figure}[t]
	\centering %[trim=left bottom right top, clip]
	\includegraphics[trim=3.3cm 11.4cm 3.6cm 11.0cm,clip,width=11cm]{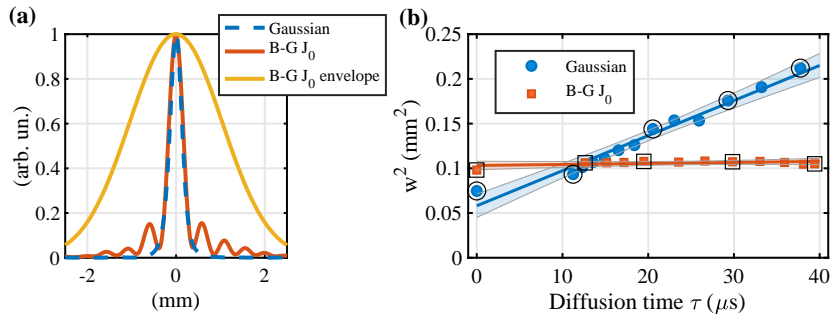}
	\caption{\label{fig:width_comparison} \small {Quantitative experimental comparison between the coherent diffusion of a Gaussian beam and a zero-order Bessel-Gauss beam. \textbf{(a)} Measured cross-sections of the input beams. Also shown is the wide Gaussian envelope of the Bessel-Gauss beam. \textbf{(b)} Squared waists radii of the Gaussian beam (blue circles) and the central lobe of the Bessel-Gauss beam (red squares) as a function of diffusion time. Values for $\tau=0$ are taken from the input beams. Solid lines are linear fits to the data; the shaded areas represent 95\% confidence intervals for the linear fit parameters. The circled data points correspond to the intensity profiles shown in Fig.~\ref{fig:Beams_table}.}}
\end{figure}

We begin by studying the coherent diffusion of Bessel-Gauss beams. We imprint on a Gaussian beam the Bessel functions $J_n$ of radial orders $n=0,1$, and 3, to serve as input probe fields, and we monitor their diffusion by recording the signal beam for different $\tau$. The results are shown in Fig.~\ref{fig:Beams_table}. As a control experiment, we input a standard Gaussian beam whose width is similar to that of the central lobe of the $J_0$ beam [see Fig.~\ref{fig:width_comparison}(a)]. Figure \ref{fig:Beams_table} also shows the intensity profiles in the Fourier plane (insets); the ring-shaped Fourier profile of the $J_0$ beam is the same for all three Bessel-Gauss beams, as we have taken care to set the same radial frequency in all of them. As clearly evident from Fig.~\ref{fig:Beams_table}, while the standard Gaussian beam expands due to diffusion, the Bessel-Gauss beams $J_0$, $J_1$, and $J_3$ are invariant to diffusion.

The invariance to diffusing in terms of the increase in waist radii is presented in Fig.~\ref{fig:width_comparison}. Figure~\ref{fig:width_comparison}(a) shows cross-sections of the input Gaussian and $J_0$ beams. We extract the waist radii in the Gaussian case by fitting a two-dimensional Gaussian function to the measured intensity profiles. In the $J_0$ case, we take the radial average of the measured pattern and extract the full width at half maximum (FWHM) of the central lobe and define the waist radius $w=\text{FWHM}/\sqrt{2\text{log}2}$. The waist radii as a function of diffusion time $\tau$ are presented in Fig.~\ref{fig:width_comparison}(b). For the Gaussian input beam, as expected, the squared waist radii grow linearly with diffusion time according to $w(\tau)^2=w_0^2+4D\tau$ \cite{FirstenbergPRL2010}. We extract $D=\SI[separate-uncertainty = true, multi-part-units=single]{9.8(12)}{\cm\squared\per\second}$ from the linear fit, which agrees with independent measurements in our setup (Appendix~\ref{appendix:exp_setup}). Conversely, the width of the Bessel-Gauss input beam remains constant, and a linear fit to the data yields an \emph{effective} diffusion constant $D'=\SI[separate-uncertainty = true, multi-part-units=single]{0.28(35)}{\cm\squared\per\second}$ that is consistent with zero. It follows that our approximation of $J_0$ is invariant to diffusion for the duration of the experiment. Similarly to the free-space propagation of Bessel-Gauss beams, where the paraxial diffraction of the Gaussian envelope governs the evolution, here the Bessel-beam approximation is valid as long as the diffusion of the Gaussian envelope is small; for the wide Gaussian envelope used in our experiment, shown in Fig.~\ref{fig:width_comparison}(a), the expected growth after $40~\mu\mathrm{s}$ of diffusion is $3.5\%$ (from $4.35~\mathrm{mm}^2$ to $4.50~\mathrm{mm}^2$). In comparison, the area of the `narrow' Gaussian beam grows almost threefold over the same duration [Fig.~\ref{fig:width_comparison}(b)]. 

\begin{figure}[t]
	\centering
	\includegraphics[width=11cm]{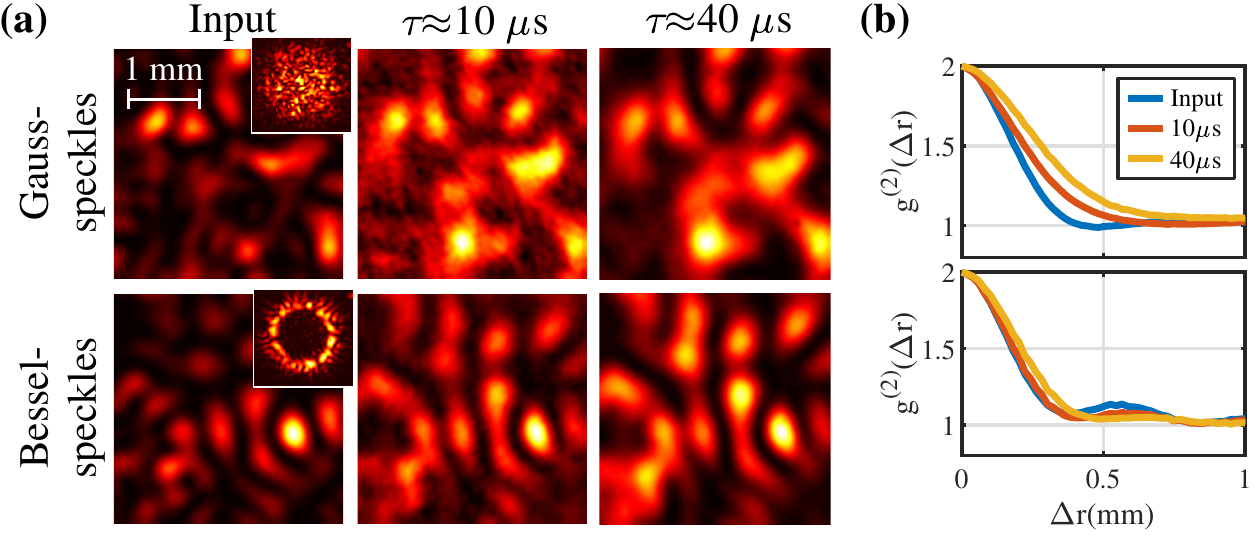}
	\caption{\label{fig:Speckle_Beams_table} \small {Coherent diffusion of Gauss speckles (top row) and Bessel Speckles (bottom row). \textbf{(a)} Left column: Measured intensity profiles of the input probe beams. Insets show the measured intensity in the far field (Fourier plane) on logarithmic scale, which has a Gaussian envelope for the Gaussian speckles and a narrow ring envelope for the Bessel Speckles. Middle and right column: retrieved speckle patterns after diffusion for $\tau\approx10~\mu\mathrm{s}$ and $\tau\approx40~\mu\mathrm{s}$. \textbf{(b)} Radial average of the autocorrelation function of the measured Gauss speckles (top) and Bessel speckles (bottom) for different diffusion time, showing a significant growth of the Gaussian speckles size due to diffusion, while the Bessel speckles largely maintain their original size.}}
\end{figure}

We now turn to diffusion experiments with two types of speckle fields, as presented in Fig.~\ref{fig:Speckle_Beams_table}(a). The input speckle patterns (left column) are shown alongside their Fourier profiles (insets). We compare a standard speckle field with a Gaussian distribution in the Fourier plane, denoted as a `Gauss speckles', to a speckle field formed by a random superposition of Bessel-Gauss beams with the same radial frequency (a ring in the Fourier plane), denoted as `Bessel speckles'. The retrieved patterns after diffusion time of 10 $\mu\mathrm{s}$ (middle) and 40 $\mu\mathrm{s}$ (right) demonstrate that the Gauss speckles grow during diffusion, while the Bessel speckles maintain their shape and size. As a quantitative analysis, we calculate the intensity autocorrelation functions of the diffusing speckle fields and present their radial average in Fig.~\ref{fig:Speckle_Beams_table}(b). For the Gauss speckles, the $1/e$ width of the autocorrelation function, which represents the average speckle grain size (and the also the coherence length \cite{Pedersen_speckles,Goodman_speckles,goodman2007speckle}) grows with diffusion time as $w^2\propto\tau$ (from 0.10 $\mathrm{mm}^2$ to 0.23 $\mathrm{mm}^2$ in 40 $\mu\mathrm{s}$). Note that this growth due to diffusion is in contrast to paraxial diffraction, for which the speckle size does not grow in the so-called deep Fresnel zone \cite{Chriki_2019_Optica}. As opposed to the Gauss speckles, the width of autocorrelation of the Bessel speckles remains nearly constant. The slow growth we observe (from 0.091 $\mathrm{mm}^2$ to 0.13 $\mathrm{mm}^2$) is due to the limited number of speckles and edge effects.

\begin{figure}[t]
	\centering
	\includegraphics[width=7cm]{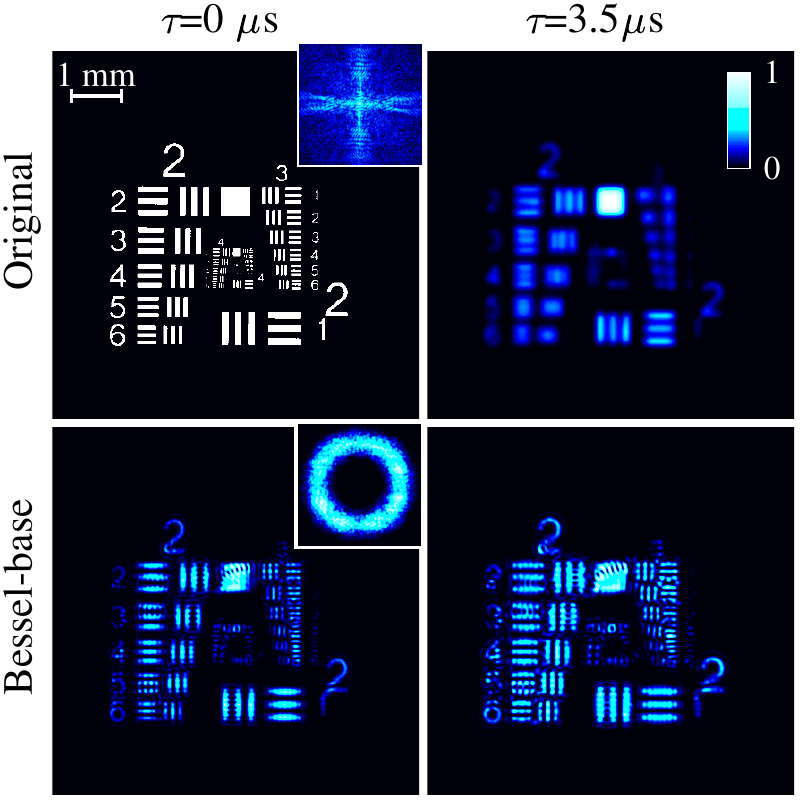}
	\caption{\label{fig:Arbitrary_beams} \small {The robustness to coherent diffusion of Bessel approximated beams. Shown are simulated intensity profiles of a USAF resolution target (top) and its Bessel approximation (bottom), before (left) and after (right) diffusion for $\tau=3.5~\mu\mathrm{s}$. Insets show the Fourier pattern of the fields on logarithmic scale, manifesting the ring-shape spatial spectra of the Bessel approximation.}}
\end{figure}

A superposition of Bessel beams can be fine tuned to approximate a desired intensity pattern, which is then rendered invariant to diffraction \cite{Lopez2010, Chriki_Shaped_beams_2018}. In analogy, such a superposition of Bessel beams would also be immune to coherent diffusion. To demonstrate this numerically, we apply a Gerchberg-Saxton iterative algorithm to approximate the intensity pattern of a USAF resolution target (in the near field), under the constraint of having a ring-shape distribution in the Fourier plane (far field). The beam we obtain is a superposition of Bessel-Gauss beams, which has an intensity pattern that approximate the USAF target, as seen in Fig.~\ref{fig:Arbitrary_beams}. This intensity pattern is encoded on the magnitude and phase along the angular coordinate of the ring the Fourier plane. Due to the reduced degrees of freedom on this ring, the obtained intensity pattern cannot be in general identical to the original (desired) pattern. The calculated evolution, presented in Fig.~ \ref{fig:Arbitrary_beams}, show that the Bessel approximation of the target is indeed invariant to diffusion over $3.5~\mu\mathrm{s}$, while the original pattern is significantly blurred.

Before concluding, we focus our attention on the role played by the phase structure of the Bessel fields, which provides an intuitive explanation for their invariance to diffusion. Consider for example the central lobe and first ring of the $J_0$ beam. Ideally, they have the same energy and opposite phases, which leads to a destructive interference that is responsible for the dark boundary between them. Under diffusion, these lobes seek to spread and to fill the dark boundary region, but owing to their opposite phases, it remains dark, and due to the equal energies, it remains stationary. Conversely, the Gaussian beam has no such supporting structure and therefore spreads when diffusing. To emphasize this point, we carry out an experiment where we effectively remove the phase from a complex-valued field when imprinting it onto the diffusing atomic coherence. To this end, we use probe and `write' control beams that share the same complex pattern, such that the imprinted field that diffuses is approximately the squared absolute value of each single beam, thus having a uniform phase [see Appendix~\ref{appendix:EIT_delay}]. Figure~\ref{fig:Squared_Beams_table} shows that the real-valued field $|J_0|^2$ and $\vert$Bessel-speckles$|^2$ spread very fast due to diffusion, in contrast to the diffusion-invariant, complex-valued fields $J_0$ and Bessel speckles, which have similar intensity patterns but alternating spatial phases.

\begin{figure}[t]
	\centering
	\includegraphics[width=10cm]{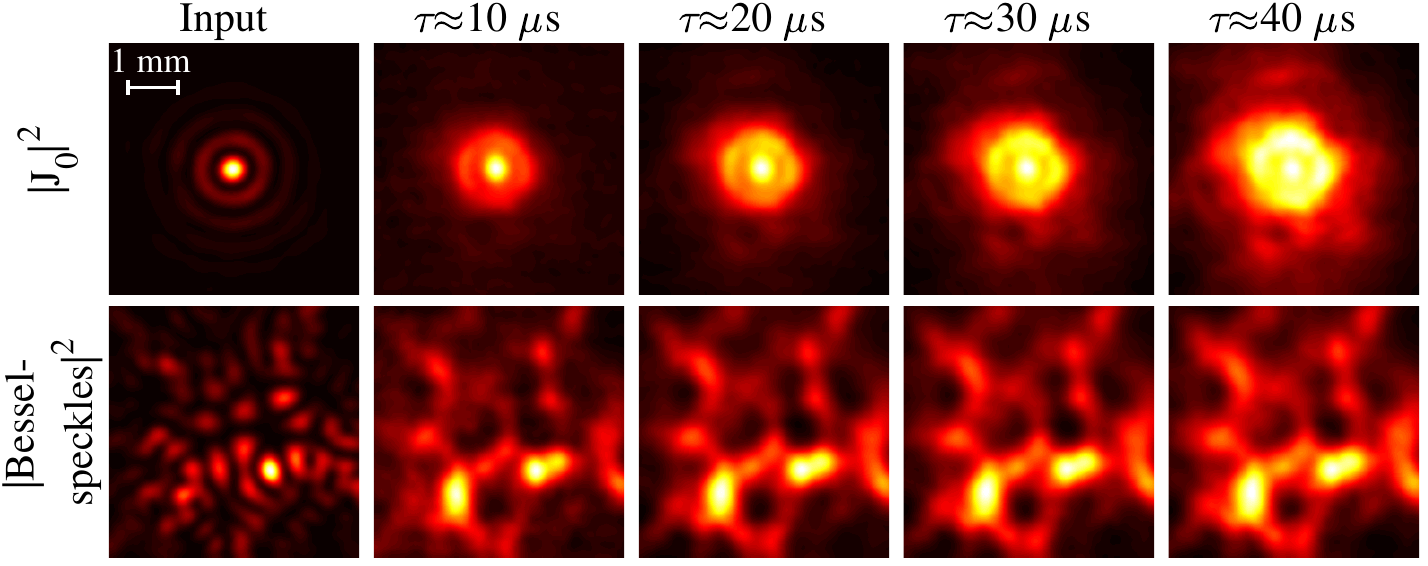}
	\caption{\label{fig:Squared_Beams_table} \small {Experimental demonstration of the importance of the phase pattern in diffusion invariance. Using the same field pattern for both the probe and the write control, we generate an input atomic coherence (left column) with a flat phase pattern and with the intensity patterns $|J_0|^2$ (top) and $|$Bessel-speckles$|^2$ (bottom). Columns labeled by $\tau\approx10$, $20$, $30$, $40~\mu\mathrm{s}$ show the measured intensity patterns after diffusion for duration $\tau$.}}
\end{figure}

\section*{Conclusion}
In summary, we study the coherent diffusion of Bessel beams and their superpositions. We experimentally compare Bessel-Gauss beams and Bessel speckles to standard Gaussian beams and speckles, quantify their evolution during diffusion, and show that the Bessel family, which is invariant to diffraction, is invariant to diffusion as well. We show, as expected, that real-valued fields with similar intensities to that of Bessel beams are not invariant to diffusion, emphasising the role played by the phase pattern of the fields. We numerically show that a superposition of Bessel beams can be used to approximate arbitrary intensity patterns and render them robust to diffusion.

\section*{Funding}
The research presented here was supported by the Israel Science Foundation (ISF) and by the Pazy foundation.

\section{Appendix}\label{appendix}

\subsection{Invariance of Bessel beams to diffusion} \label{appendix:bessel_invariance}
Briefly, invariance of Bessel beams to diffusion (and to diffraction) can be understood by considering their Fourier structure. Diffusion acts as a low-pass filter on the field, attenuating its components depending on their spatial frequency and irrespective of their direction.
A Bessel field is confined to a thin circle in Fourier space, thus comprising a single spatial frequency. Therefore all of its frequency components are equally attenuated when it diffuses (and they acquire equal phase when it diffracts). The field thus remains unchanged up to a uniform attenuation (and up to a uniform phase shift in diffraction). 

In more details, consider the diffusion equation in real space $(\partial/\partial t-D\nabla^2_\bot)\psi(\mathbf{r},t)=0$, where $\psi(\mathbf{r},t)$ is a two-dimensional complex field. This equation is written in Fourier space as $(\partial/\partial t+Dq^2)\tilde{\psi}(\mathbf{q},t)=0$. Its solution is $\tilde{\psi}(\mathbf{q},t)=\tilde{\psi}(\mathbf{q},0)\exp(-Dq^2t)$, where $\tilde{\psi}(\mathbf{q},0)$ is the initial field distribution in Fourier space. The term $\exp(-Dq^2t)$ with a real constant $D$ is the diffusion propagator; it is radially symmetric, depending only on $q=|\mathbf{q}|$. A Bessel beam of order $n$ is given by $\psi_n(\mathbf{r})=J_n(q_0r)\exp(in\theta)$, where $\mathbf{r}=(r,\theta)$, and its Fourier transform $\tilde{\psi}(q,\theta_q,z=0)=(2\pi/q)(i)^{-n}\exp(in\theta_q)\delta(q-q_0)$ forms a circle of radius $q_0$ around $\mathbf{q}=0$ \cite{Baddour2011}. When multiplying by the diffusion propagator in Fourier space, the circle is preserved up to the global attenuation $\exp(-Dq^2t)$, and consequently the intensity pattern in real space is conserved. The attenuation is independent of the order $n$ of the Bessel beam, and thus the invariance holds for any superposition $\sum_{n}^{}a_nJ_n(q_0r)\exp(in\theta)$, where $a_n$ are complex coefficients.

In the experiments and simulations, we use an approximation of Bessel beams, so-called Bessel-Gauss beams. These beams are ideal Bessel beams with an added Gaussian envelope in the near field \cite{Bessel_Gauss}. In Fourier space (far field), they form a ring with a finite width, as seen in Figs.~\ref{fig:Beams_table},~\ref{fig:Speckle_Beams_table}, and~\ref{fig:Arbitrary_beams}). They are therefore nearly invariant to diffusion up to a time set by the width of the ring, which is inversely proportional to the width of the Gaussian envelope.

\subsection{Experimental arrangement}
\label{appendix:exp_setup}

An amplified 795-nm diode laser with (one photon) linewidth of \SI{1}{\mega\hertz} is split into three beams. The `write' and `read' control beams overlap spatially in the area of the vapor cell, but are separated by an angle of $\theta\approx\SI{10}{\milli\radian}$. The 'probe' beam is modulated at $\sim\SI{6.8}{\giga\hertz}$ and reflected from the SLM, which sets its complex transverse profile. A blazed grating is imprinted on the SLM on top of the desired beam pattern in order to separate the shaped field from the undiffracted part. The two-photon detuning $\Delta_{\text{2p}}$ is scanned by varying the modulation frequency of the probe by at most $\pm\SI{15}{\kilo\hertz}$. The probe is oriented such that it propagates along the `write' control beam inside the vapor cell. The three beams generate a fourth `signal' beam along the path of the `read' control. All fields are circularly polarized; the `write' control and the probe are $\sigma^+$, whereas the 'read' control and the generated signal are $\sigma^-$. The control beams are nearly top-hat beams with diameter $\gtrsim\SI{8}{\milli\meter}$, except for the experiments in Fig.~\ref{fig:Squared_Beams_table}, where the `write' control is given the shape of the probe (then both the probe and the `write' control exit the same optical fiber and together are spatially modulated using the SLM). After the cell, the generated signal is spatially separated from the `write' control and from the probe and filtered from the 'read' control using a pair of Fabry-P\'{e}rot etalons. A pair of lenses is used to image the signal onto a CCD camera.

We use \Rb vapor with \SI{10}{\torr} of $\text{N}_2$ buffer gas, heated to \SI{65}{\celsius} and placed in a \SI{7.5}{\cm} long cell. The cell is held inside a three-layered shield to isolate it from the external magnetic field, and a weak \SI{50}{\milli\gauss} longitudinal magnetic field generated with Helmholtz coils shifts the spectator Zeeman states in the ground level $5S_{1/2}$ away from the Raman resonance, thus assuring that only the states $\ket{g}=\ket{5S_{1/2}; F=1,2; m=0}$ participate in the process. In the excited level $5P_{1/2}$, both hyperfine levels, namely the four state $\ket{e_1}=\ket{5P_{1/2}; F'=1,2; m=1}$ and $\ket{e_2}=\ket{5P_{1/2}, F'=1,2; m=-1}$, participate in the process. The diffusion constant in the cell $D=\SI[separate-uncertainty = true, multi-part-units=single]{9.7(5)}{cm\squared\per\second}$ is measured independently utilizing standard light storage experiments~\cite{vudyasetu2008storage,FirstenbergPRL2010, smartsev2017continuous} and agrees with the calculated values~\cite{ma2009modification, ishikawa2000diffusion} (for further details refer to Supplementary Material of Ref.~\cite{Chriki_2019_Optica}).

%=================================================================

\subsection{Diffusion and induced diffraction in EIT four-wave mixing medium}
\label{appendix:EIT_delay}
We rotate the basis of excited states and define $\ket{\pm}=\left(\ket{e_2}\pm\ket{e_1}\right)/\sqrt{2}$ and the corresponding normal modes $E_+$ and $E_-$. Accordingly, we express the normal modes immediately at the exit of the cell as $E_\pm^{out}=g_\pm E_\pm^{in}$, where $g_-=e^{-S}$ describes regular (one-photon) absorption, and $g_+ = e^{-S\left(1-f\right)}$ describes EIT. Here $S$ and $f$ are the complex Lorentzian profiles associated with the one and two photon resonances,
%%%%%
\begin{equation}
\label{eq:define_SandF}
S = d\frac{\gamma_{\text{1p}}}{\gamma_{\text{1p}}-i\Delta_{\text{1p}}},\qquad \mathrm{and} \qquad
f = \eta_{\mathrm{act}}\frac{\Gamma}{\gamma_{\text{2p}}+\Gamma-i\Delta_{\text{2p}}},
\end{equation}
%%%%%
where $2d$ is the resonant optical depth for the probe, $\Delta_{\text{1p}}$ and $\Delta_{\text{2p}}$ are the one-photon and two-photon frequency detunings, $\gamma_{\text{1p}}$ and $\gamma_{\text{2p}}$ are the corresponding decoherence rates, and $\Gamma$ denotes the power broadening $\Gamma=\Omega^2/\left(\gamma_{\text{1p}}-i\Delta_{\text{1p}}\right)$ with $\Omega$ the Rabi frequency of the control beams. The prefactor $0\leq\eta_{\mathrm{act}}\leq1$ is the fraction of atoms that populate the $m=0$ Zeeman state in the lower hyperfine manifold. More details are given in Ref.~\cite{smartsev2017continuous}.

For a uniform incoming probe field $E_\text{in}$, the generated signal $E_s$ in the limit of weak EIT $|Sf|\ll 1$ is given by
\begin{equation}
\frac{E_s}{E_\text{in}} = \frac{1}{2}\left(g_+ - g_-\right)\approx{\frac{1}{2}Sfe^{-S}},
\label{eq:generated1}
\end{equation}
Substituting $S$ and $f$ and taking the derivative with respect to frequency, we find that the generated signal has a group delay of
\begin{equation}
\tau_\text{tot}=\frac{\partial}{\partial\Delta_{\text{2p}}}\left[\log\left(\frac{g_+-g_-}{2}\right)\right] =\frac{\gamma_{\text{2p}}+\Gamma-i\Delta_{\text{2p}}}{\left(\gamma_{\text{2p}}+\Gamma\right)^2+\left(\Delta_{\text{2p}}\right)^2}.
\label{eq:diffusion_time1}
\end{equation}
It follows that $\tau_\text{tot} = \tau+i\tau_\text{dr}$ has both real and imaginary components. The real part $\tau$ corresponds to real diffusion, while the imaginary part $\tau_\text{dr}$ corresponds to motional-induced diffraction. In the experiment, we vary the diffusion time $\tau$ by changing the two photon detuning $\Delta_{\text{2p}}$. Notice that for a given $\gamma_{\text{2p}}+\Gamma =\mathrm{const}$, the maximal diffusion time $\tau$ is obtained for $\Delta_{\text{2p}}=0$. Accordingly, we denote the maximal diffusion time $\tau_\infty\equiv1/(\gamma_{\text{2p}}+\Gamma)$, resulting in 
\begin{equation}
\frac{\tau_\text{tot}}{\tau_\infty}= \frac{1-i\Delta_{\text{2p}}\tau_\infty}{1+\left(\Delta_{\text{2p}}\tau_\infty\right)^2}.
\label{eq:diffusion_time2}
\end{equation}
In our experiments, $\tau_\infty\approx\SI{40}{\micro\second}$.

To estimate the effect of the induced diffraction, we consider the evolution of a Gaussian beam of initial waist $w_0$. In analogy to the Rayleigh range in optics, we define a Rayleigh duration $\tau_\text{R}$ after which the the beam expands by a factor of $\sqrt{2}$ due to diffusion. The width of the beam after some diffusion time $\tau$ is simply 
\begin{equation}
w(\tau)^2=w_0^2+4D\tau,
\end{equation}
and therefore $\tau_\text{R}=w_0^2/(4D)$. The induced diffraction becomes significant if the duration $\tau_\text{dr}$ is large when compared to the effect of diffusion,
\begin{equation}
|\tau_\text{dr}|>\tau_\text{R}+\tau.
\end{equation}
Plugging Eq.~(\ref{eq:diffusion_time2}) and solving for $\Delta_{\text{2p}}\tau_\infty$, we find for our experimental parameters that considerable induced diffraction is expected for $140>|\tau_\infty\Delta_{\text{2p}}|>3.3$. We can therefore safely neglect the induced diffraction in the regime $3>\tau_\infty\Delta_{\text{2p}}>0$ studied in the manuscript.

%==========================
Using the Fourier transformation $\Tilde{E}(\mathbf{q}) = \int \frac{d^2\mathbf{r}}{2\pi} E(\mathbf{r})e^{-i\mathbf{q}\cdot\mathbf{r}}$ for the transverse coordinates $\mathbf{r}=(x,y)$ and under the assumptions of weak EIT and confined spatial frequencies $q^2=|\mathbf{q}|^2\ll|\gamma_{2p} + \Gamma|/D$, it can be shown that \cite{smartsev2017continuous}
%%%%%
\begin{equation}
\label{eq:DiffusionInFourier}
\Tilde{E}_s(\mathbf{q})\propto\Tilde{E}_\text{in}(\mathbf{q})e^{-D\tau q^2},
\end{equation}
where $\Tilde{E}_\text{in}$ and $\Tilde{E}_s$ are the fields of the incoming probe and the generated signal in Fourier space. 
The propagator $e^{-D\tau q^2}$ in Fourier space implies diffusion in real space. It follows %from Eq.~(\ref{eq:DiffusionInFourier}) 
that a structured probe beam in our system continuously generates a signal which underwent diffusion for an effective temporal duration $\tau$. 

%==================================

To study diffusion-invariant fields, we spatially structure the complex probe field ${E}_\text{in}(\mathbf{r})$ in the transverse plane. When the `write` control field is wide and uniform, the atomic (dark state) coherence acquires that complex structure $\sigma(\mathbf{r})\propto{E}_\text{in}(\mathbf{r})$. An exception is the the experiment presented in Fig.~\ref{fig:Squared_Beams_table}, where the `write' control has the same spatial shape as the input probe $\Omega(\mathbf{r})\propto {E}_\text{in}(\mathbf{r})$ [$\Omega(\mathbf{r})$ is the control Rabi frequency], and the atomic coherence is real valued,
\begin{equation}
\sigma(\mathbf{r})\propto
\frac
{E_\text{in}(\mathbf{r}) \Omega(\mathbf{r})^*}
{|\Omega(\mathbf{r})|^2/\Gamma+\gamma_{2p}}\propto \frac{|{E}_\text{in}(\mathbf{r})|^2}
{|\Omega(\mathbf{r})|^2/\Gamma+\gamma_{2p}}.
\end{equation}
As $|\Omega(\mathbf{r})|^2/\Gamma$ cannot be neglected with respect to $\gamma_{2p}$, we end up with real-valued atomic coherence shaped as a regulated approximation of $|{E}_\text{in}(\mathbf{r})|^2$.

%=======================================

%%%%%%%%%%%%%%%%%%%%%%% References %%%%%%%%%%%%%%%%%%%%%%%%%

\bibliography{Diffusion_free}
\bibliographystyle{abbrv}
\end{document}